\documentclass{article}

\usepackage{arxiv}

\usepackage[utf8]{inputenc} 
\usepackage[T1]{fontenc}    
\usepackage{hyperref}       
\usepackage{url}            
\usepackage{booktabs}       
\usepackage{amsfonts}       
\usepackage{nicefrac}       
\usepackage{microtype}      
\usepackage{lipsum}		
\usepackage{graphicx}
\usepackage{natbib}
\usepackage{doi}

\title{Immersive In Situ Visualizations for Monitoring Architectural-Scale Multiuser MR Experiences}

\author{ 
\href{https://orcid.org/0000-0002-3671-1619}
{\includegraphics[scale=0.06]{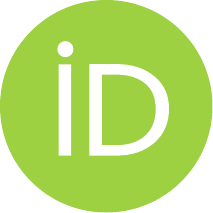} \hspace{1mm}Zhongyuan Yu}
\thanks{This project is funded by the European Union and co-financed from tax revenues on the basis of the budget adopted by the Saxon State Parliament (Project No. 100690214).} \\
Immersive Experience Lab \\
Technische Universität Dresden\\
Helmholtzstr. 10, 01069 Dresden \\
\texttt{zhongyuan.yu@tu-dresden.de} \\
\And
\href{https://orcid.org/0000-0003-4954-4896}
{\includegraphics[scale=0.06]{orcid.pdf} \hspace{1mm}Daniel Zeidler} \\
Immersive Experience Lab \\
Technische Universität Dresden\\
Helmholtzstr. 10, 01069 Dresden \\
\texttt{daniel.zeidler@tu-dresden.de} \\
\And
\href{https://orcid.org/0000-0002-5718-7043}
{\includegraphics[scale=0.06]{orcid.pdf} \hspace{1mm}Krishnan Chandran} \\
Immersive Experience Lab \\
Technische Universität Dresden\\
Helmholtzstr. 10, 01069 Dresden \\
\texttt{krishnan.chandran@tu-dresden.de} \\
\And
\href{https://orcid.org/0000-0002-9268-4854}
{\includegraphics[scale=0.06]{orcid.pdf} \hspace{1mm}Lars Engeln} \\
Immersive Experience Lab \\
Technische Universität Dresden\\
Helmholtzstr. 10, 01069 Dresden \\
\texttt{lars.engeln@tu-dresden.de} \\
\And
\href{https://orcid.org/0009-0005-5978-5513}
{\includegraphics[scale=0.06]{orcid.pdf} \hspace{1mm}Kelsang Mende} \\
Immersive Experience Lab \\
Technische Universität Dresden\\
Helmholtzstr. 10, 01069 Dresden \\
\texttt{kelsang.mende@tu-dresden.de} \\
\And
\href{https://orcid.org/0000-0002-8923-6284}
{\includegraphics[scale=0.06]{orcid.pdf} \hspace{1mm}Matthew McGinity} \\
Immersive Experience Lab \\
Technische Universität Dresden\\
Helmholtzstr. 10, 01069 Dresden \\
\texttt{matthew.mcGinity@tu-dresden.de} \\
}

\hypersetup{
pdftitle={A template for the arxiv style},
pdfsubject={q-bio.NC, q-bio.QM},
pdfauthor={David S.~Hippocampus, Elias D.~Striatum},
pdfkeywords={First keyword, Second keyword, More},
}

\begin{document}
\maketitle

\begin{abstract}
Mixed reality (MR) environments provide great value in displaying 3D virtual content.
Systems facilitating co-located multiuser MR (Co-MUMR) experiences allow multiple users to co-present in a shared immersive virtual environment with natural locomotion.
They can be used to support a broad spectrum of applications such as immersive presentations, public exhibitions, psychological experiments, etc. 
However, based on our experiences in delivering Co-MUMR experiences in large architectures and our reflections, we noticed that the crucial challenge for hosts to ensure the visitors’ quality of experience is their lack of insight into the real-time information regarding visitor engagement, device performance, and system events. 
This work facilitates the display of such information by introducing immersive in situ visualizations.
\end{abstract}


\section{Introduction}

Over recent years, multiuser MR (MUMR) systems, in which groups of people share a single, coherent world composed of virtual and real elements, have been shown to hold great promise, with applications spanning education, gaming, and data analysis.
By enabling direct perception of both the physical world and fellow users, MUMR fosters collective, collaborative behavior, and contributes to overcoming isolation experienced with pure VR \cite{guo_breaking_2024}. 

Recently, low-cost mobile HMDs with video see-through capabilities, such as Meta Quest 3s, allow for seamless and flexible adjustment in the integration between real and virtual environments across the entire reality-virtuality continuum \cite{continuum}.
Besides, with the emergence of robust markerless inside-out tracking and on-device computing, such devices can now be effectively used in large spaces without depending on any external tracking or computing infrastructure. 
Hence MUMR systems built on such devices become highly scalable, allowing visitors to explore expansive virtual scenarios by naturally walking through large architectural spaces, making them ideally suited for creating immersive presentations.

We use the term \textit{Co-MUMR} to refer to \textit{Co-Located  Multiuser Mixed Reality} experiences. 
Further, we define a Co-MUMR experience to be ``architectural scale'' when it enables viewers to navigate freely within spaces that exceed the dimensions of typical room-sized environments and involves structures such as hallways, staircases, mezzanines, elevators, escalators, etc \cite{viewr}. 
In our lab, we have developed and deployed many \textit{Architectural-scale Co-MUMR} experiences, in artistic, educational, museum, research studies or prototypes, gaming, and psychotherapy domains, gaining many hundreds of hours of experience with hundreds of subjects. 
In all cases, we observe roles are typically differentiated into ``host'' and ``visitors''. Depending on the context, the ``host'' may be a tour guide, teacher, instructor, moderator, experiment coordinator, or support technician, while the ``visitors'' may be museum or exhibition visitors, trainees or students, or trial subjects, for example. 
The ``host'' typically has experience with the system, and is responsible for ensuring a smooth experience for everyone involved. 
Sometimes multiple hosts are active, or the roles change dynamically.

Very often, the hosts themselves are wearing an HMD, because they have an active role as guiding experts in the experience. 
In this case, they face the challenge of performing their ``in-experience'' diegetic role (e.g. teacher), while simultaneously performing the role of host. 
However, even in cases where the host has no active diegetic role, our experience demonstrates that the performing of the duties of a host can be greatly enhanced when the host themselves is wearing an HMD. 

In this paper, we therefore explore the possibilities and investigate the usefulness of in situ mixed reality visualizations for hosting architectural-scale Co-MUMR experiences. 
We facilitate the display of real-time visitor and device information with immersive in situ visualizations, empowering hosts with direct information access.
The visualized information serves as an initial step towards enabling informed decision-making, assisting in resolving technical issues, or potentially offering targeted assistance and explanations to visitors during the experience.

Our main contributions are:
(1) The design of immersive in situ visualizations for monitoring visitor engagement and system performance towards aiding hosts in enhancing visitor experiences.
(2) The implementation of the proposed in situ visualizations in an operational system.

\begin{figure}
    \centering
    \includegraphics[width=\linewidth]{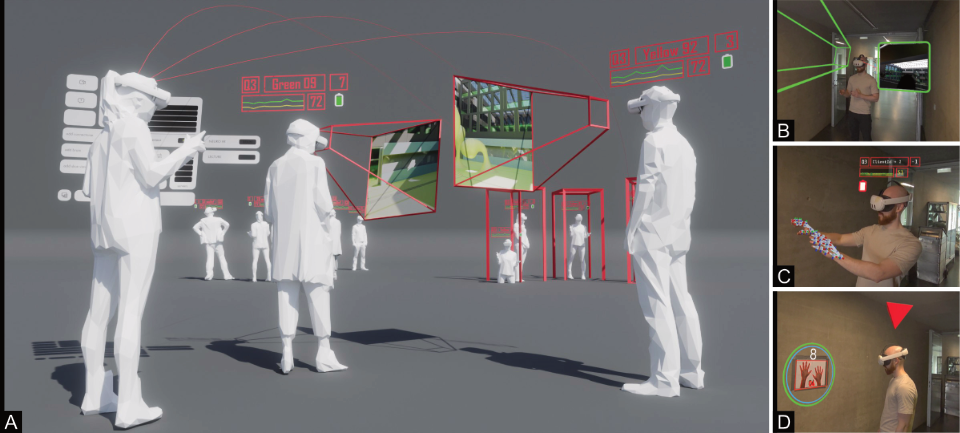}
    \caption{Exemplary immersive in situ visualizations.
  (A) A sketch of a collection of in situ visualizations helping hosts access real-time information regarding the multiuser MR experience.
  (B) Visitor engagement visualization displays their rendered view by side.
  (C) Embedded visualization of real-time device metrics, including rendering framerate and battery status.
  (D) Visualization of past calibration events with a color-coded circular-shaped chart around a calibration station. }
    \label{fig:enter-label}
\end{figure}

\section{Related Work} 
A few approaches have investigated in situ visualizations of visitors' engagement, such as view directions, directly in the immersive environment. 
For instance, Wang et al. proposed a system to enhance transparency and integrative transition between the realities of HMD users sharing the same physical environment \cite{SliceofLight}.
It allows guests to observe all other HMD users’ interactions contextualized in their own MR environment.
However, this approach primarily aims at enabling guests to perceive each other's perspectives without distinguishing between hosts and visitors and it is not specifically tailored to address the challenge of understanding visitor engagement in large architectural spaces.
Thanyadit et al. developed a system to enable instructors to observe and guide students in the context of VR classrooms \cite{ObserVAR}. They proposed immersive visualizations to enhance the instructor's ability to monitor students' activities and focus. 
However, the system primarily caters to educational settings where the virtual environment for each visitor is consistently synchronized, which is considered as a ``guided'' experience, visualizations to offer insights into visitor engagement in an ``explorative'' setting are not included in their work.

\section{Information Expectations and Design Considerations}

\begin{figure}
\centering
  \includegraphics[width=\linewidth]{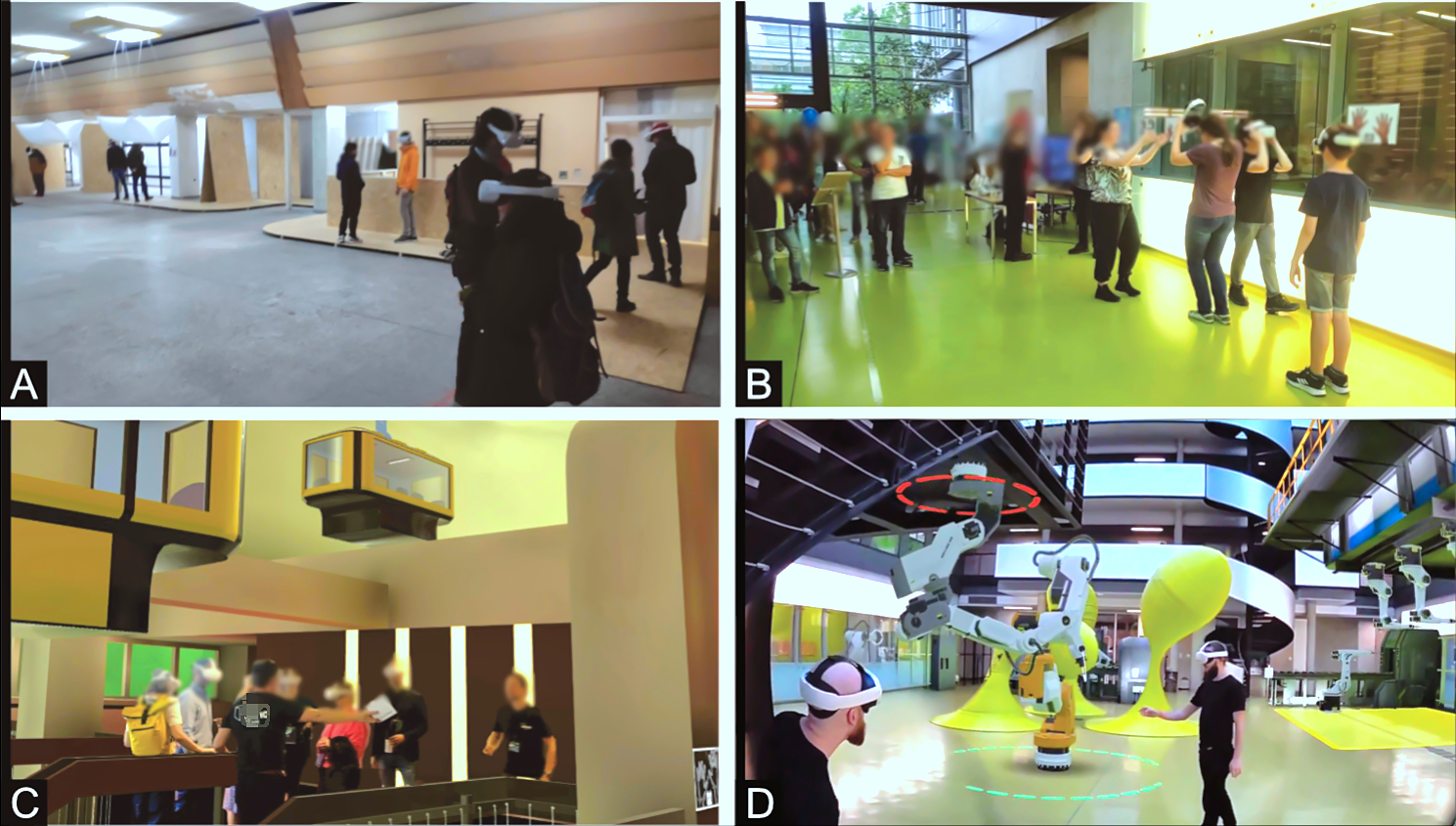}
  \caption{
    Photos and screenshots of our previously organized and hosted Co-MUMR experiences. 
  }
  \label{fig:Co-MUMR Background}
\end{figure}

\subsection{Co-MUMR Hosting Experiences} 
\label{FourExperiences}
Throughout the past years, our lab has developed and exhibited many Co-MUMR experiences, in a variety of public and research domains, ranging from public exhibitions in museums to multi-user educational experiences and large-scale MUMR scientific studies in psychology and psychotherapy (as shown in Fig.~\ref{fig:Co-MUMR Background}). Collectively, this constitutes many hundreds of hours of hands-on MUMR experience with many hundreds of visitors. 
We present four such cases here: 

\paragraph{E1: Hybrid Spaces.}
In February 2023, we developed and hosted a Co-MUMR experience that blended real and virtual elements, transforming a 30m x 20m space into an immersive multi-layered parkour environment.
This explorative experience spanned four days, with 180 visitors, each on average staying 30 minutes in MR (see Fig. \ref{fig:Co-MUMR Background} (A)).
The hosts were responsible for introducing the concept to participants, aiding the tracking alignment, and resolving issues for unexpected disruptions. Hosts did not wear MR headsets themselves.

\paragraph{E2: Virtual Leap of Faith.}
Around the same period, we conducted a second Co-MUMR experience, augmenting a four-story building with a large space (20m x 30m) by superimposing it with an artistic version of the space. 
This Co-MUMR experience spanned one day, engaging around 100 visitors. Each session accommodated a minimum of two participants, managed by three individuals
(see Fig. \ref{fig:Co-MUMR Background} (B)).

\paragraph{E3: Future Mobility.}
In mid-2023, we organized an additional Co-MUMR experience in the foyer and on the terrace of the second floor (around 15m x 5m each), with a focus on future climate-neutral mobility concepts. 
This one-day event drew the interest of approximately 50 visitors (see Fig. \ref{fig:Co-MUMR Background} (C)).

\paragraph{E4: Robot Lab.}
In early 2024, we developed and hosted a Co-MUMR robot lab experience in a large foyer space (20m x 30m) with around 20 participants, aimed to illustrate the concepts of human-robot interaction (see Fig. \ref{fig:Co-MUMR Background} (D)).

\subsection{Challenges and Expectations} 
Given the novelty and the limited access to Co-MUMR experiences at the time of writing, very few people have adequate experience in hosting such experiences except our colleagues.  
Thus in this work, we conducted meta-reflection sessions with four experienced hosts within our team (including colleagues and authors, all hosted at least two Co-MUMR experiences before) based on those previous hosting experiences (E1-4) and collected the challenges and expectations. 
The sessions are organized as interviews and each session lasts around 60 min. 
During the interview, we encouraged the hosts to recall and enumerate all challenges encountered during their hosting and their anticipated solutions uninterrupted \cite{CollaborativeReflection}. 
We then filtered the ones that could potentially be addressed through immersive in situ visualizations.
The findings are in the following.

During E1, without information on the visitors' rendered view, it was difficult for hosts to determine whether the visitors were \textit{viewing the intended content} or mistakenly entered another scene or opened a wrong app.
When assisting visitors with issues, the hosts always started by communicating verbally, \textit{inquiring about what the visitors were seeing}, sometimes the hosts even had to take over visitors' headsets to understand their perspective.
There was a need for hosts to understand visitors' perspectives efficiently and effectively. 
%
Besides, as described above, the visitor flow was extensive. The hosts found it challenging to provide headsets with sufficient \textit{battery life}. There is a risk of visitors' headset batteries depleting unexpectedly. 
Moreover, due to the complex nature of Co-MUMR systems and visitor interactions, hosts encountered unexpected technical issues, 
including network offline events, calibration failures, and frame drops.
There is a need to show \textit{system events} in their occurring places to help identify issues.
In addition to the visualization of system events, for diagnosing the issues, it is necessary to delve deeper by examining \textit{detailed device performance history} before and after the event time point.

In E2, hosting an artistic experience, the hosts expected the virtual content to be delivered optimally with the best possible device performance.
This requires hosts to remain attentive to technical factors as the experience runs, including \textit{rendering frame rate, CPU/GPU usage, battery levels, positional tracking, hand tracking, network strength, bandwidth, etc} and be ready to assist.

Co-MUMR experience E3 was held in a multi-story building with obstacles. The hosts noticed that visitors sometimes spread out, making it difficult to \textit{track where they went}, let alone assist them. 
Besides, the network connection was not stable -- visitors might suddenly go offline due to unexpected reasons such as going out of range, low battery life, or pressing the wrong button.
There is a need to be aware of such \textit{device offline events} to aid in debugging network issues. 

During E4, when demonstrating the manipulation of virtual robot arms, hosts noticed that it wasn't clear if visitors fully understood the steps taken and whether the visitors could see the intended actions being performed. 
Questions like \textit{``Are you seeing what I am seeing?''} was frequently observed.

\subsection{Key Information for Hosts} 
Based on the above reflections, we observed that hosts face a recurring challenge related to the hindered awareness of the entire operational context and they require real-time information to provide the necessary assistance.
We summarize the expected information as follows:

\label{InfoVisExp}
\paragraph{I1: Information Regarding Visitor Engagement.} \label{I1}
There is a need for hosts to understand how visitors are engaging with and enjoying the Co-MUMR experience. This includes: \textit{What are visitors seeing and doing? Where are they?}
Thus, it is classified as visitors' perspectives (E1, E4) and locations (E3).

\paragraph{I2: Information Regarding Device Performance.}  \label{I2}
Besides, there is a need to know the real-time status of device performance and make sure they are working optimally. 
This could be classified as the devices' general performance (including rendering frame rate,
CPU/GPU usage, 3D position, etc) (E1, E2), hand tracking performance (E2), and networking performance (E2).
Device performance history for inspecting system events is also desired based on E1. 

\paragraph{I3: Information Regarding Real-time Event.}  \label{I3} 
We note that based on the experience in E1 and E3, there is a need for hosts to be aware of system events, including networking events (E3) and calibration events (E1) to help debug system functions.  

\subsection{Design Considerations}

In this work, we aim to deliver the expected information (I1, I2, I3) to hosts through visualizations, enhancing their awareness and effectiveness when monitoring the visualizations.
Through monitoring, hosts are empowered with great information access and improved awareness of issues, which could lead to potential actions such as active visitor flow guidance to improve the visiting experience.

\paragraph{Immersive In Situ Visualization.}
Based on our experience, in many cases, the host must join the immersive experience to share his spatial movements with the visitors and be in the same virtual reality while hosting.
While it is also possible to display the visitor and system information on an additional 2D display, we target immersive in situ mixed-reality visualizations in this work given the following reasons:
1) The spatial and dynamic nature of the desired information (I1, I2, I3) is well-suited to be displayed spatially in mixed reality. 
For example, the visitors' movement is originally in 3D and would move around the space constantly. 
2) Wearing the same mixed-reality headsets as visitors allows hosts to experience the same virtual environment, enabling a better understanding of the visitors' perspectives.
3) Embedded visualizations \cite{Wesley_Embedded_Data_Representations} in mixed reality allow users to effortlessly access desired information (I1, I2, I3) by simply approaching the relevant subject, thus eliminating the need for context switching.

\paragraph{Adaptivity.} 
\label{VisAdaptivity}
By offering visualization alternatives, hosts can transform the visualizations into various forms, enabling them to meet their monitoring needs. 
We designed visualizations to be adaptive in large architectural spaces, ranging from \textit{subject-centric:} visualizations placed embedded by the side of the relevant subjects (visitors, devices) to \textit{host-centric}: visualizations projected in front of the host.
For \textit{host-centric} visualizations, we enable hosts to adjust the visual displays to either eye or floor level.

\section{Visualization Design and Implementation}
Here, we designed a series of immersive in situ visualizations and refined the visual style of the visualizations iteratively with an experienced visual designer.  
The visualizations are classified according to the desired information in Sec. \ref{InfoVisExp} with Sec. \ref{VisitorEngagement} designed for displaying I1 (information regarding visitor engagement), Sec. \ref{HMD performance} for I2 (information regarding device performance), and Sec. \ref{SystemEvent} for I3 (information regarding real-time event).
At runtime, the visualizations can be configured separately and integrated on demand with great flexibility with an MR GUI panel. 

Note that in the following figures, the items shown in grayscale indicate scene-specific or background components, while the colorized elements highlight the proposed visualizations. 
The reddish color is used solely to emphasize the visualization concepts presented in this work. In practical applications, the color scheme is customizable to suit specific needs or preferences.


\subsection{Visitor Engagement Visualization} \label{VisitorEngagement}

Based on the analysis of desired visitor information in Sec. \ref{I1}, in this section, we describe the proposed visualizations to show visitors' perspective with a detailed view (visitors’ rendered view in Sec. \ref{Visitors' Rendered View}) and coarse overview (real-time view frustum in Sec. \ref{Visitors' Rendered View}), 
location of individual visitors (Sec. \ref{VisitorLocator}) and a group of visitors as an overview (Sec. \ref{GroupLocator}).

\begin{figure*}
\centering
  \includegraphics[width=\linewidth]{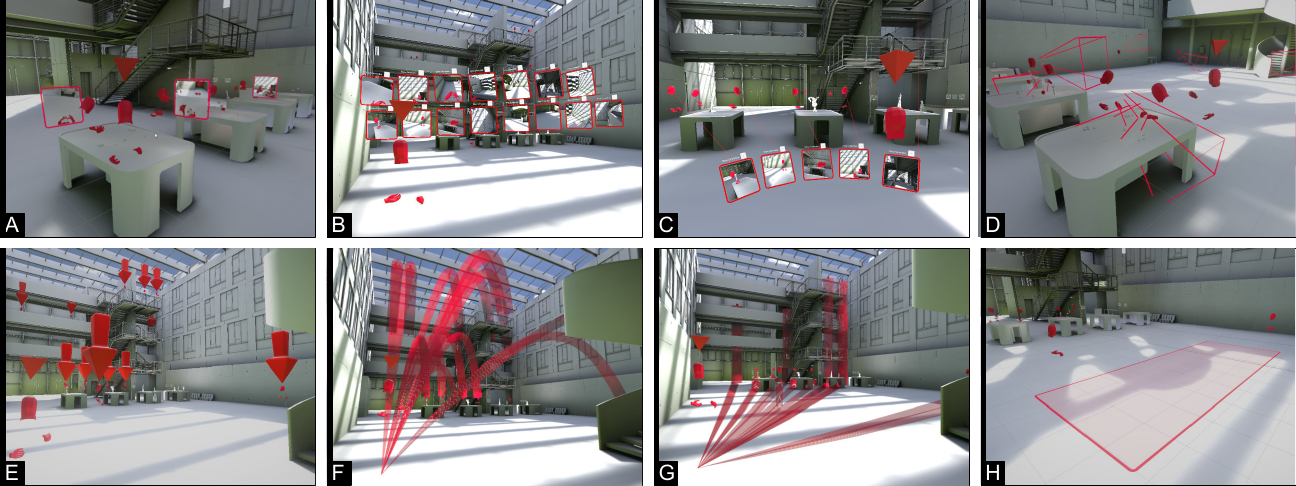}
  \caption{
    The screenshot of the proposed visitor engagement visualizations for displaying
    visitors' detailed rendered views placed in space embedded by the side of visitors (A), projected in front of the host (B and C), and a coarse view of visitors' perspective with view frustums (D). 
    Spatial locations of individual visitors with embedded giant arrows (E), visual links from the host to visitors (F and G), and a coarse view of the spatial locations of a group of visitors (H).
    }
  \label{fig:visitor engagement}
\end{figure*}

\subsubsection{Visitors' Rendered View and View Frustum} \label{Visitors' Rendered View}
Showing visitors' rendered view serves as an intuitive approach for hosts to understand their engagement (I1). However, built-in view monitoring functions on most MR headsets trend to stream the views to a website or a mobile application, providing minimal support for monitoring the view directly from another MR device.   
For this, we developed a customized code block to handle the data transmission and support the display of visitors' rendered views directly within the host's headset. 
When rendering in an immersive space, instead of the conventional approach of placing all views on a large panel, we support embedded visualization by showing a view frustum for each visitor and subsequently texturing the visitor's perspective directly onto the forefront of this frustum.
However, given the dynamic nature of the visitors' viewing directions, view frustums could move constantly. 
Thus, we suggest attaching the view adjacent to visitors and keeping the view oriented to the host to show the detailed visitors' rendered view (as shown in Fig. \ref{fig:visitor engagement} (A)) while using the view frustum as an overview of visitor engagement (Fig. \ref{fig:visitor engagement} (D)). 

In large spaces, however, the attached views could be less visible when visitors are distant from the host. 
To enhance the visibility of the views, we propose to project visitors' views onto a hemispherical surface near the host for host-centric monitoring.  
Nevertheless, when dealing with numerous visitors, their views can overlap. 
To address this, we also support the automatic organization of the views into a grid, resulting in a body-centric array of views \cite{ens_spatial_2017} (see Fig. \ref{fig:visitor engagement} (B) and (C)).

\subsubsection{Locating Individual Visitors} \label{VisitorLocator}
In Co-MUMR experiences, identifying a visitor's location is relatively simple when they are near the host, but it becomes challenging when they interact with scene elements lie beyond the host's immediate field of view (similar issue reported in \cite{Arradar,LabelingOutofViewObjects}).
One option for indicating visitors' locations in architectural Co-MUMR experiences is to strengthen the visual appearance of visitors with embedded visual cues.
The ghosted views proposed in \cite{ghostedviews} would be helpful, but when monitoring in large spaces, a larger and stronger visual cue is desired. 
Therefore, a giant arrow is designed to appear above each visitor (as shown in Fig. \ref{fig:visitor engagement} (E)).
Another option is to show a visual link (similar to \cite{prouzeau_visual_2019}) between visitors and the host.
Here we utilize spatial curves.
We implemented these curves to originate from a point fixed in front of the host and extend toward the visitors' heads.
These curves gradually widen as the distance from the host increases, enhancing its visibility when viewed from a distance.
The maximum width is configurable. 
Besides, these curves are marked with animated arrow patterns for directionality. 
The animation could serve as a visual guide to help hosts follow the spatial curves.
From the host's perspective, each curve uniquely identifies a visitor, enabling the host to physically approach the visitor by following the curve's trajectory (as shown in Fig. \ref{fig:visitor engagement} (F)).
To reduce visual clutter, this spatial curve can be configured to be flattened on the floor (as shown in Fig. \ref{fig:visitor engagement} (G)), and visitors within a certain distance in the viewing range can be excluded.

\subsubsection{Overview of Visitor Spatial Locations} \label{GroupLocator}

To help provide an overview of the visitors' location, we further propose an area indicator on the ground to highlight the area and boundary of a group of visitors. 
We implemented the area indicator in a square shape, it dynamically shifts and adjusts its dimensions in real time in response to the visitors' movements.
A screenshot is shown in Fig. \ref{fig:visitor engagement} (H).

\subsection{Device Performance Visualization} \label{HMD performance}

\begin{figure*}
\centering
  \includegraphics[width=\linewidth]{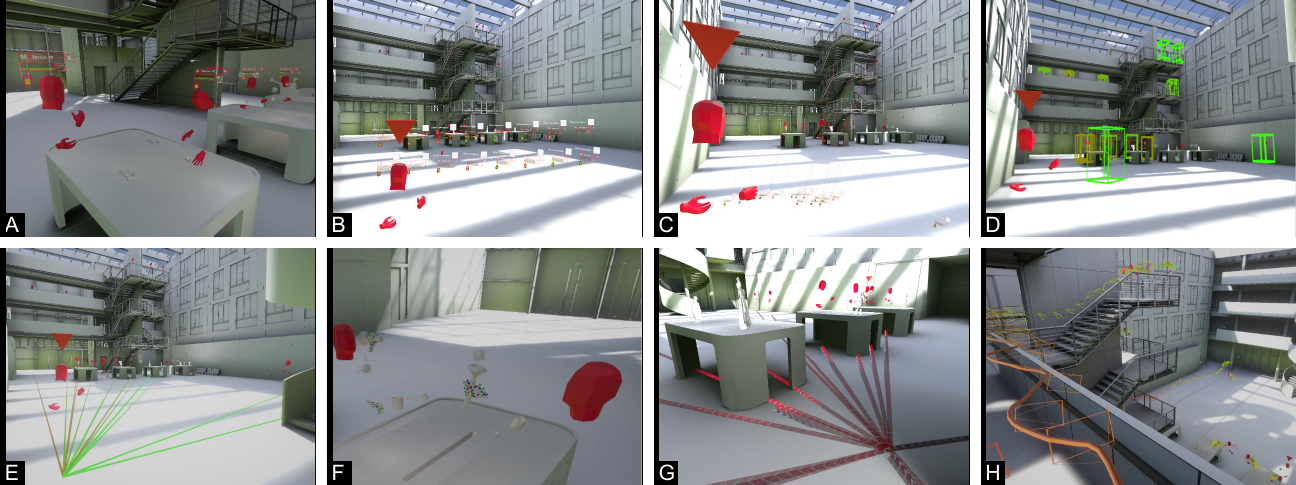}
  \caption{
    Screenshots of the visualizations for showing detailed performance metrics with information panel placed by side of individual HMDs (A), projected in front of the host (B and C).
    Coarse overview of the performance by encoding into the color of visitor bounding boxes (D) and visual links between visitors and the host (E).
    Hand tracking performance in (F) and networking performance in (G).
    Device performance history alongside its spatial movement trajectory in (H).
  }
  \label{fig:hmd performance vis}
\end{figure*}

Even when visitors are smoothly engaging with the Co-MUMR system, due to the complexity and immature nature of the technique, technical issues can readily happen at any time. 
Hosts should be aware of such issues immediately and be prepared to assist in resolving any potential problems.
During the experiences, hosts can roughly know a headset's tracking status when a passthrough bubble around visitors is shown (the case in \cite{viewr}). 
However, details regarding the performance of individual devices, especially the performance of the HMDs, including the rendering efficiency, battery level, and tracking accuracy, remain obscured.

\subsubsection{Embedded Information Panel} \label{EmbInfoPanel}
The effectiveness of embedded in situ visualizations has been demonstrated in the field of augmented reality \cite{RagRug}. 
However, little work has been dedicated to the visualization of HMD information in co-located multiuser MR scenarios.
An intuitive solution is to show information panels. 
This panel is designed to be oriented towards the host at all times and display the status of visitors' headsets to help visualize I2 (similar to the panel for displaying the AR device's status in \cite{cavallo_cave-ar_2019}).
In the current implementation, we offer two options to show the information panels and allow hosts to configure them freely to adapt to various situations under various situations when hosting in large architectural spaces.
One option is to render the information panel embedded atop visitors' individual headsets as shown in Fig. \ref{fig:hmd performance vis} (A). 
Another option is to render the panel projected to the host in front at eye level or on the floor (see Fig. \ref{fig:hmd performance vis} (B) and (C)). 
The embedded panels are mainly designed for monitoring near visitors while the projected panels are designed for monitoring at a distance.
In our system, real-time performance data on individual HMDs such as the battery life and rendering FPS, are collected on the visitors' headsets and then transmitted to hosts in real time for visualization.

\subsubsection{Performance Overview with Visitor Bounding Box and Color-encoded Links} \label{AvatarBoundingBox} \label{FPSLine}
To provide an overview of the panel-based textual representation when conveying HMD performance information, we explore a stronger visual cue to help hosts identify the performance issues.   
One option is to show a bounding box-like frame (to avoid blocking vision) and encode performance metrics in the color around visitors. 
We currently implemented a color-coded bounding box around visitors as an example. It is shown in Fig. \ref{fig:hmd performance vis} (D).  
Another option is to show a visual link between the host and the visitors. 
An example is shown in Fig. \ref{fig:hmd performance vis} (E).
We also refer to this visualization as ``FPS Lines'' in this specific example.
The color of the bounding box and the visual link transitions from green to red and reflects the real-time frame rate value. 
Moreover, we note that additional glyphs or potential metaphors could be considered as alternatives and integrated into the system in the future.

\subsubsection{Head and Hand Tracking Visualization} \label{HandTrackingVis}
In this work, we implemented the real-time data transmission and visualization of live head and hand poses.
We utilized a polygonal virtual head to display the head pose. 
For hand poses, we support the visualization of detailed joint positions and coordinate axes, to aid in debugging hand skeleton tracking (Fig. \ref{fig:hmd performance vis} (F)).
This visualization becomes especially useful when superimposing it to the video see-through, which enables direct comparison between the tracked hand poses in the system and the real hand movement. 
In the future, hand tracking issues could be detected and aggregated over time into a visualization, aiding hosts in diagnosing and debugging the system further.

\subsubsection{Network Traffic Visualization} \label{NetworkTrafficVis}
For certain experiences with media content streams in the network, the visualization of the network bandwidth as well as incoming and outgoing network packages would help hosts identify the network bottlenecks and other potential network issues. 
Similar to \cite{WiFiVis}, this work visualizes network data transmission within a 3D spatial context to help further show I2. 
Distinctively, we employ 3D curves that have specific start and end points corresponding to the physical locations of devices in the environment in situ.
Those curves show arrow-like patterns in bi-directional with certain widths and directions, and are animated. 
The direction indicates the direction of data transmission, the animation speed reflects the network latency and the curve width depicts the network bandwidth (see Fig. \ref{fig:hmd performance vis} (G)).
We note that for this visualization, we mainly focus on the visualization design and prototyping, the real network data is not currently integrated.   
%

\subsubsection{Device Performance History} \label{PerformanceAroundEvents}
We propose a visualization of \textit{HMD performance history} combined with its \textit{spatial movement history}, represented as multivariate encoded trajectories in space, assisting hosts in diagnosing issues related to device performance.
The spatial movement history comprises both positional movement and rotational movement (relevant to view direction history), and is visualized as a spatial trajectory with mini-frustums placed at specific intervals to indicate the visitor's view direction over time, similar to the camera trajectories described in \cite{dynascape}. 

At the same time, the HMD performance history, which includes metrics such as battery life, frame rate, and CPU usage over time, is traced and can be visualized alongside the spatial movement trajectories.
They can be encoded into the texture or mapped onto specific geometries along the trajectory.
An example is shown in Fig. \ref{fig:hmd performance vis} (H), where each trajectory represents an individual HMD, and the color of each segment of the trajectory reflects the corresponding rendering framerate.
In the current implementation, the movement and performance data of all HMDs have been traced since the system started. 
At a certain time, hosts can request and visualize the cumulative data up to that point. 
%

\subsection{System Event Visualization} \label{SystemEvent}
 
\begin{figure}
\centering
  \includegraphics[width=\linewidth]{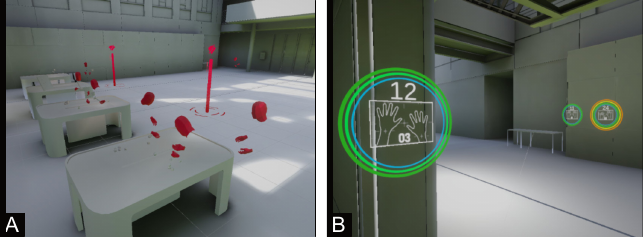}
  \caption{
    Visualizations for showing system events: offline events (A) and calibration events (B).
  }
  \label{fig:system event vis}
\end{figure}

\subsubsection{Network and Tracking Loss Event} \label{NetworkEventVis} 

In the system, we monitor internet connection and visitors' positional tracking quality, emitting events when there are status changes.
This visualization helps hosts recognize sudden offline events and identify their spatial patterns.
A screenshot of two sudden offline events is shown in Fig. \ref{fig:system event vis} (A).  
%

\subsubsection{Calibration Event} \label{CalibrationVis}
As depicted in Fig. \ref{fig:system event vis} (B), we accumulate the count of calibration events and show this data by encircling the calibration stations with color-coded circles. The color transitions from blue to red from inner to outer circles.   
The frequency of visitor calibration at a specific station is directly proportional to the number of circles displayed around that station.
%
This color coding aims to help hosts quickly discern the importance of a calibration station from a distance with a glance.
The insights from this visualization can be used to help guide hosts in optimizing the layout of calibration stations or virtual installations. 

\section{Technical Details}

We implemented our system based on an open-source framework \cite{viewr}. We customized and optimized the system's capabilities, focusing on networking, user interface, session recording, and immersive rendering. 
We replaced the normcore plugin in the original framework with Unity Netcode. This adjustment facilitates high-speed data transmission in a local network and grants complete flexibility in managing the networking logic.
Our system mainly targets Meta Quest devices including Quest 2, Quest 3, and Quest Pro. 
They are highly portable and mobile, can be used to support Co-MUMR on large architectural spaces with extensive visitors. 
All visualizations proposed in this work are implemented and optimized for real-time rendering in standalone mode on the device (without PC). 
Upon acceptance, we plan to make the source code publicly available to support further research in this domain. 

\begin{figure}
    \centering
    \includegraphics[width=\linewidth]{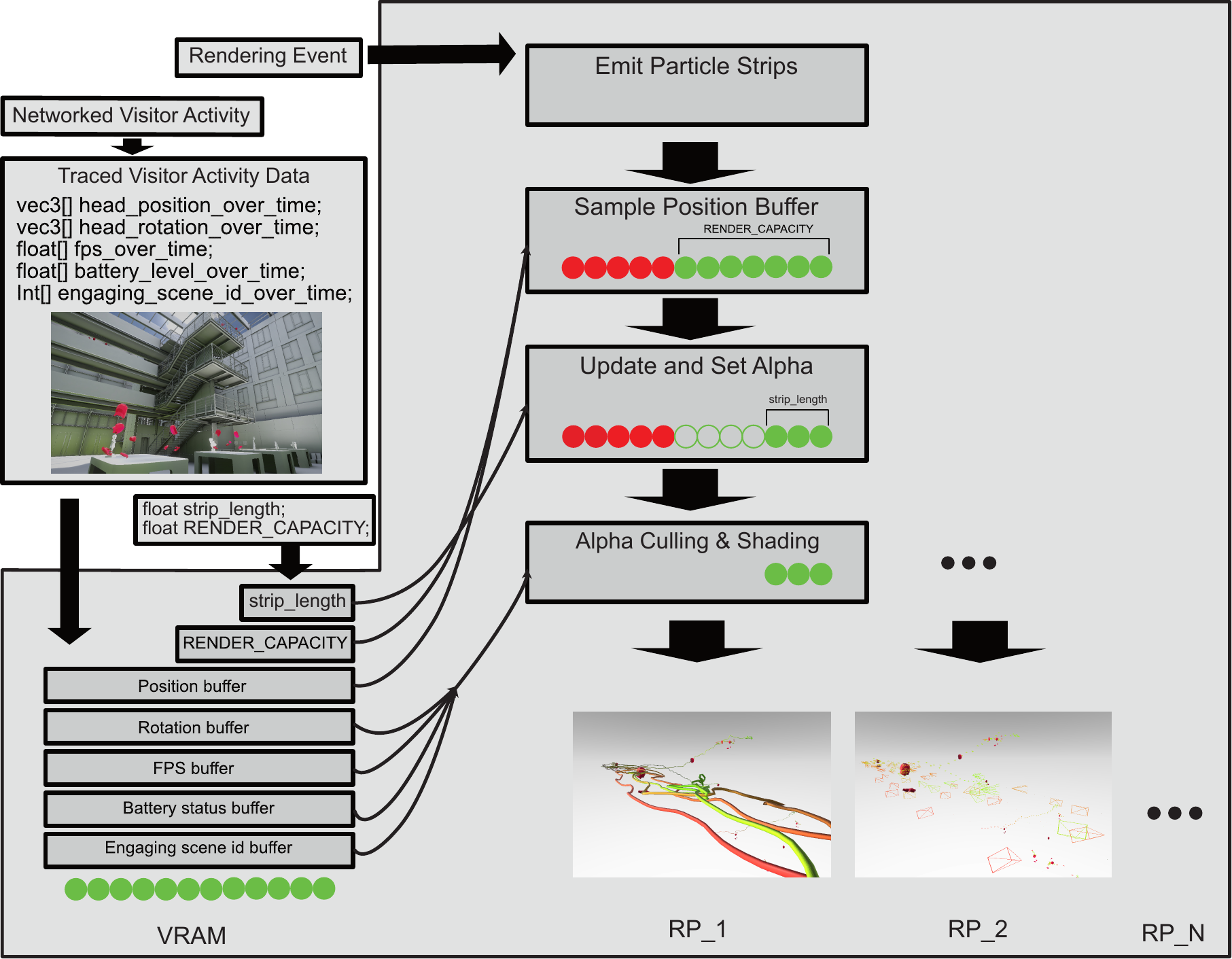}
    \caption{Data flow for displaying device performance around events with spatial trajectories. 
    The visitor activity data is traced and truncated according to a real-time configurable variable with alpha culling.
    The data buffers in video RAM (VRAM) (shown in the bottom left corner) can be used by multiple render paths.}
    \label{fig:data flow}
\end{figure}

The 3D spatial curves proposed in this work (such as the ones proposed in section \ref{VisitorLocator}, \ref{NetworkTrafficVis}, and \ref{PerformanceAroundEvents}) are rendered by Unity's “VFX Graph” with the "Particle Strips" block.
They are visualized as spatial ribbons, with interpolated normals to achieve a solid visual appearance. 
Furthermore, Unity's "VFX Graph" supports a separate texturing step with a "Shader Graph", which offers simplicity and flexibility in rendering patterns on the trajectories similar to \cite{OnTubeRendering}. This function is highly beneficial for multivariate information visualization tasks.
For the rendering of device performance history as 3D trajectories (Sec. \ref{PerformanceAroundEvents}), the traced data samples undergo filtering based on spatial and directional variances before being uploaded for rendering, the rendering data flow is illustrated in Fig. \ref{fig:data flow}

\section{Conclusion and Future Work}

In this paper, we proposed immersive in situ visualizations to help convey visitor information to hosts in real time in architectural-scale multi-user experiences. 

We developed methods to reduce visual clutter by preventing overlapping visualizations when the number of visual elements becomes excessive.
Further mechanisms could be introduced to address visual clutter in scenarios involving a large number of visitors.
This work designed visualizations to assist in visualizing visitor engagements. However, fully understanding these engagements poses a broader and more complex challenge. 
Future designs could focus on providing deeper insights into the nature and nuances of engagement information.
With an increasing number of visitors and larger spatial environments, hosting becomes an even more challenging task. 
Future designs and mechanisms could be developed to support more efficient and effective hosting in such complex scenarios.
For instance, a prioritization mechanism based on proximity and the severity of visitors' needs could be introduced to enhance the responsiveness and decision-making capabilities of the hosts.

We tested the system function by delivering an exemplary Co-MUMR puppet exhibition.
A complete evaluation study of the system's functionality and the effectiveness of the visualizations could be conducted in future work.
Additionally, an investigation into balancing the reduction of visual clutter against the risk of increased information loss could be conducted to guide further refinement.

\bibliographystyle{unsrtnat}
\bibliography{template}  






\end{document}